\documentclass[a4paper,12pt]{article}
\usepackage{amssymb,latexsym}

\catcode`\@=11

\def\section{\@startsection{section}{1}{\z@}{3.5ex plus 1ex minus
 .2ex}{2.3ex plus .2ex}{\bf}}

\def\thesubsection{\arabic{section}\arabic{subsection}}
\renewcommand{\subsection}[1]{\addtocounter{subsection}{1}
\vspace{2.5mm}\par\noindent {\it \thesubsection . #1}\par
\vspace{0.5mm} } \catcode`\@=12

\newcommand{\SO}{\mathrm{SO}}
\newcommand{\GL}{\mathrm{GL}}
\newcommand{\SL}{\mathrm{SL}}
\newcommand{\SU}{\mathrm{SU}}

\newcommand{\U}{\mathrm{U}}
\newcommand{\E}{\mathrm{E}}

\mathchardef\varGamma="0100 \mathchardef\varDelta="0101
\mathchardef\varTheta="0102 \mathchardef\varLambda="0103
\mathchardef\varXi="0104 \mathchardef\varPi="0105
\mathchardef\varSigma="0106 \mathchardef\varUpsilon="0107
\mathchardef\varPhi="0108 \mathchardef\varPsi="0109
\mathchardef\varOmega="010A

\def\bfone{\relax{\rm 1\kern-.35em 1}}

\DeclareFontFamily{U}{rsf}{} \DeclareFontShape{U}{rsf}{m}{n}{
  <5> <6> rsfs5 <7> <8> <9> rsfs7 <10-> rsfs10}{}
\DeclareMathAlphabet\Scr{U}{rsf}{m}{n}

\makeatletter \@addtoreset{equation}{section} \makeatother

\begin{document}

\begin{titlepage}

\thispagestyle{empty}

\begin{flushright}
\hfill{CERN-PH/TH/2004-054} \\
\hfill{hep-th/?}
\end{flushright}

\vspace{35pt}

\begin{center}{ \LARGE{\bf
Homogeneous Special Manifolds, Orientifolds and Solvable
Coordinates.}}

\vspace{60pt}

{\bf R. D'Auria $^\star$, S. Ferrara $^{\dagger\ddag\sharp}$ and
M. Trigiante $^\star $}

\vspace{15pt}

$\dagger${\it  CERN, Theory Division, CH 1211 Geneva 23,
Switzerland. }\\ $\ddag${\it INFN, Laboratori Nucleari di
Frascati, Italy.}\\
$\sharp$ {\it University of California, Los Angeles, USA}

 \vspace{15pt}

$\star${\it Dipartimento di Fisica, Politecnico di Torino \\
C.so Duca degli Abruzzi, 24, I-10129 Torino, and\\
Istituto Nazionale di Fisica Nucleare, Sezione di Torino, \\
Italy}

\vspace{50pt}

{ABSTRACT}

\end{center}

\medskip
We discuss some geometrical properties of the underlying $N=2$
geometry which encompasses some low--energy aspects of $N=1$
orientifolds as well as four dimensional $N=2$ Lagrangians
including bulk and open string moduli.\par In the former case we
illustrate how properly defined involutions allow to define $N=1$
K\"ahler subspaces of special quaternionic manifolds. In the
latter case we show that the full shift symmetry of the brane
coordinates, which is abelian in the rigid limit, is partially
distorted by bulk fields to a nilpotent algebra.

\end{titlepage}
\section{Introduction}
In superstring theory it is possible to construct models where
bulk and brane degrees of freedom build a supersymmetric theory in
lower dimensions with some number  $N$ of supersymmetries left
unbroken. Of particular interest are models with $N=4,2,1$
supersymmetry where gauge and matter degrees of freedom are
present. Models with $N=4$ supersymmetry are quite restricted.
They can be obtained, for example, by compactification of Type IIB
theory on a $T^6$ orientifold where the bulk sector describes
gravity plus six vector multiplets and the brane sector describes
some additional gauge multiplets. The Lagrangians of these models
(in absence of fluxes) will differ by the way duality symmetries
act on the vector fields. Recalling that the $\sigma$--model
manifold is
\begin{eqnarray}
\frac{\SU(1,1)}{\U(1)}\times
\frac{\SO(6,6+n)}{\SO(6)\times\SO(6+n) }\,,
\end{eqnarray}
the difference come from the way the various vector fields
transform under some different subgroups of $\SU(1,1)\times
\SO(6,6+n)$. The simplest examples are two extreme cases, the
$T^6/\mathbb{Z}_2$ orientifolds with $D3$--branes, and Type I
superstring, i.e. a Type IIB on a $T^6$--orientifold with
$D9$--branes present. In the latter case the full $ \SO(6,6+n)$
acts on the $12+n$ vector potentials, while in the former case
only a $\GL(6,\mathbb{R})\times \SO(n)$ acts linearly on the
vector potentials. Moreover $\SU(1,1)$ acts linearly on the $12$
bulk vectors of the former while it acts as an electric--magnetic
duality on the 12 bulk vectors of the latter. A consequence of
this situation is that the 15 shift symmetries of the
$\SO(6,6)/(\SO(6)\times \SO(6))$ manifold do not act on the 12
bulk  vector potentials in the $T^6/\mathbb{Z}_2$--orientifolds
while they act on some potentials in the Type I case.\par For
$N=2,1$ theories this state of affairs is even more involved
because different $Dp$--branes can occur at the same time and so
the geometry of  bulk and different brane moduli must be explaned.
Moreover the  compactification manifold can involve curved spaces
such as $K3$ or Calabi--Yau manifolds. In the present paper we
consider some geometrical aspects which play a role in the
supergravity description of these models.\par The paper is
organized as follows: in section 2 we give two $N=4$ examples,
namely the $T^6/\mathbb{Z}_2$--orientifold with $D3$--branes, and
Type I on $T^6$. We show that the bulk sector of these theories
can be obtained using two different Lie algebra involutions which
truncate the original $N=8$ solvable Lie algebra of
$\E_{7(7)}/\SU(8)$ to a $ \frac{\SU(1,1)}{\U(1)}\times
\frac{\SO(6,6)}{\SO(6)\times\SO(6) }$ submanifold.\par In section
3 we consider, in the $N=2$ context, a class of homogeneous
quaternionic manifolds which allow to obtain, using suitable
involutions defined by Cartan isometry generators, different $N=1$
spaces upon truncation. One of these spaces is related to the
c--map \cite{cfg,fs} and another to a ``dual'' K\"ahler space
which occur in Calabi--Yau orientifolds \cite{bghl,bbhl,ggjl,dft}.
They differ by the fact that different NS and R--R scalars are
present depending on the particular involutions used.
\par In section 4 we analyze a class of rank 3 homogeneous Very
Special Geometries that are relevant for $N=2$ theories in the
presence of $D3$ and $D7$--branes\footnote{The relation between
$D3/D7$ Special Geometry and certain homogeneous manifolds was
brought to our attention by A. Van Proeyen.}. We describe the
relation between the solvable Lie algebra coordinates and the
holomorphic special coordinates and their relation to the brane
coordinates as obtained from the combined Born--Infeld,
Chern--Simons action. We analyze the full set of shift symmetries
in the $2n_3+2 n_7$ brane coordinates of which only $n_3+ n_7$ are
abelian, as in the rigid limit of the Born--Infeld action. The
other $n_3+ n_7$ shift symmetries do not commute with the previous
ones  due to the presence of Chern--Simons terms which must be
added to the Born--Infeld action. However the full B--I+C--S
action is further corrected to reproduce the metric of the
predicted Special Geometry. The microscopic origin of these
corrections is related to sypersymmetry and gauge anomaly
counterterms which arise when compactifying  the $D=6$ parent
theory down to $D=5$ or $D=4$ \cite{fms,abfpt}.
\par These results are relevant to understand the relation of
these formulations with the computations made in terms of
Born--Infeld actions. The relations between different coordinates
is also relevant for cosmological applications of supersymmetric
models with branes, where shift symmetries in the brane
coordinates play an important role \cite{kp,wy,l,hk}.\par In
section 5 we recall the six--dimensional origin of the
prepotential and its possible generalizations.\par Finally in
 Appendix \ref{appendiceA} we recall the geometry of the dual
 $N=1$ K\"ahler manifold and give the explicit expression for its
 Riemann tensor.
\section{$N=8$ involutions and two $N=4$ examples}
\label{truncation} Let us start reviewing the construction of the
two ungauged $N=4$ models describing the bulk sector of Type IIB
on $T^6/\mathbb{Z}_2$ orientifold, in the presence of $D3$ and
Type IIB on $T^6$ orientifold in the presence of $D9$--branes
respectively (the latter model, as mentioned in the introduction,
being equivalent to the bulk sector of Type I on $T^6$). The first
case was analyzed in detail in \cite{kst,fp,dfv}. In this section
only we shall label by $a,b$ the Dirichlet directions and by $i,j$
the Neumann directions of the internal torus. In the presence of
$D3$--branes transverse to $T^6$ all the compact directions will
be Dirichlet and the orientifold projection has the form $\Omega
I_6 (-)^{F_L}$, where $\Omega$ is the world--sheet parity and
$I_6$ is the space parity acting on the torus coordinates. The
surviving scalar fields (in the bulk sector) are
$\phi,\,g_{ab},\,C_{abcd},\, C_{(0)}$ while the vector fields are
$C_{a\mu},\,B_{a\mu}$. As far as the latter case is concerned, the
orientifold projection (consistent with residual $N=4$ in the
presence of $D9$--branes) consists of the world--sheet parity
$\Omega$ only. The surviving scalar fields (in the bulk sector)
are $\phi,\,g_{ij},\,C_{ij},\, C_{\mu\nu}^*$ ($C_{\mu\nu}^*$ being
the scalar field dual to the four--dimensional tensor field
$C_{\mu\nu}$) while the vector fields are $C_{i\mu},\,G_{\mu}^i$.
These two models can be obtained as truncations of the $N=8$
theory describing the low--energy limit of Type IIB on $T^6$ in
absence of fluxes. At the supergravity level the orientifold
operations indeed are realized by  involutions on the solvable
algebra generating the $N=8$ scalar manifold $\E_{7(7)}/\SU(8)$
and on the vector fields. Later on we shall briefly review the
solvable Lie algebra formalism. Consider the scaling symmetry
defined by a Cartan generator $h$ with respect to which the scalar
fields transform with an integer grading $k$: $\Phi_k\rightarrow
e^{k\lambda}\, \Phi_k$. This is still a symmetry of the Lagrangian
even if we extend the scaling parameter to the complex plane and
set $\lambda=i\pi$. This scaling transformation now takes the
form:
\begin{eqnarray}
\Phi_k&\rightarrow & (-)^k \Phi_k\,\,\Rightarrow \,\,
\cases{\Phi_{2\ell}\rightarrow
\Phi_{2\ell}\cr\Phi_{2\ell+1}\rightarrow -\Phi_{2\ell+1}}\,.
\end{eqnarray}
This implies that setting $\Phi_{2\ell+1}=0$ defines a consistent
truncation of the scalar sector since the corresponding solvable
Lie algebra closes. This truncation can be generalized to
4--dimensional $p$--forms by setting to zero all the fields which
are odd with respect to the involution $(-)^{h+p}$. As a
consequence of this the scalar fields and rank 2 tensor fields of
the truncated theory should have even grading with respect to $h$
while vector fields should have an odd grading. This recipe for
truncation can be extended to four dimensional \emph{gauged}
supergravities corresponding to string theory compactifications in
the presence of fluxes. From the supergravity point of view a flux
in the microscopic setting is described by an \emph{embedding
matrix} which is a tensor transforming covariantly with respect to
the global symmetry group ${\Scr G}_e$ of the ungauged lagrangian
and which defines how the gauge group ${\Scr G}$ is embedded
inside ${\Scr G}_e$. \emph{The gauged Lagrangian is still globally
invariant under ${\Scr G}_e$ provided that besides the fields also
 fluxes are transformed}. In particular also fluxes, in the
low--energy description, have a well defined grading under any
dilation symmetry of the ungauged Lagrangian. Since fluxes are
0--form field strengths, from the four dimensional point of view
they can be associated with $p=-1$ forms and therefore our recipe
for truncation would require restricting to those fluxes which are
even with respect to $(-1)^{h-1}$. This is consistent with the
requirement that the minimal couplings $(Flux)\times A_\mu\times
\partial^\mu \Phi$ and the non abelian terms $(Flux)\times
A_\mu\times A_\nu\times F^{\mu\nu}$ be even under $(-1)^h$.
Summarizing: in a gauged supergravity describing the low--energy
limit of a superstring compactification with fluxes, a Cartan
generator $h$ of a symmetry of the Lagrangian such that
$(-)^{2h}=+1$ on the bosonic sector, defines a consistent
truncation of the theory if we restrict the fields to the
following gradings:
\begin{eqnarray}
\mbox{ truncation in
4-D}&:&\cases{\mbox{scalars}\,\,\,(-1)^h=+1\cr\mbox{vectors}\,\,\,(-1)^h=-1\cr
\mbox{2-forms}\,\,\,(-1)^h=+1\cr\mbox{fluxes/embedding--matrix}\,\,\,(-1)^h=-1\cr\cr}\,.
\end{eqnarray}
 Let us now briefly recall the main facts about
 the solvable Lie algebra description of homogeneous
scalar manifolds in supergravity. We refer the reader to
\cite{adfft,dwst} where the relevant conventions are fixed. This
representation consists in describing a homogeneous scalar
manifold ${\cal M}$ as a Lie group generated by a solvable Lie
algebra $Solv({\cal M})$, whose parameters are the scalar fields
of the theory:
\begin{eqnarray}
{\cal M}&=&\exp[Solv({\cal M})]\,,
\end{eqnarray}
All homogeneous manifolds which are relevant to supergravity admit
such representation.
 In the $N=8$
model derived from Type IIB toroidal compactification, the
solvable Lie algebra generating the scalar manifold can be
described as follows: {\small \begin{eqnarray}
Solv_7&=&Solv\left(\frac{\E_{7(7)}}{\SU(8)}\right)= Solv_2+
Solv_6+{\bf (2,15)}_{+1}+{\bf (1,15^\prime)}_{+2}+{\bf
(2,1)}_{+3}\,, \nonumber\\Solv_2&=&
Solv\left(\frac{\SL(2,\mathbb{R})_{{\rm
IIB}}}{\SO(2)}\right)=\{\phi H^\prime+C_{(0)}\, t\}\,,\nonumber\\
Solv_6&=&Solv\left(\frac{\GL(6,\mathbb{R})}{\SO(6)}\right)=\{\sum_{\Lambda=1}^6
\log (g_{\Lambda\Lambda}) H_\Lambda+\sum_{\Lambda\neq \Sigma}
g_{\Lambda \Sigma}\, t^\Lambda{}_\Sigma\}\,,\nonumber\\
{\bf (2,15)}_{+1}&=&\{B_{\Lambda\Sigma\tau}\,
t^{\Lambda\Sigma\tau}\}\,\,;\,\,\,\,{\bf
(1,15^\prime)}_{+2}=\{\epsilon^{\Lambda_1\dots\Lambda_6}\,C_{\Lambda_1\dots
\Lambda_4}\, t_{\Lambda_5\Lambda_6}\}\,,\nonumber\\{\bf
(2,1)}_{+3} &=&\{D_\tau \, t^\tau\}\,,\label{IIBsolv}
\end{eqnarray}}
where $\Lambda, \Sigma=1,\dots, 6$ label the directions of the
internal torus, $\SL(2,\mathbb{R})_{{\rm IIB}}$ is the duality
group of the ten dimensional Type IIB theory, $\tau,\,\sigma=1,2$
are indices of the doublet of $\SL(2,\mathbb{R})_{{\rm IIB}}$ so
that
$\{B_{\Lambda\Sigma\tau}\}=\{C_{\Lambda\Sigma},\,B_{\Lambda\Sigma}\}$
and $\{D_\tau\}=\{B_{\mu\nu}^*,\,C_{\mu\nu}^*\}$. The group
$\GL(6,\mathbb{R})$ acts transitively on the $T^6$ metric moduli
$g_{\Lambda\Sigma}$ and the representations in (\ref{IIBsolv})
refer to the group $\SL(2,\mathbb{R})_{{\rm IIB}}\times
\GL(6,\mathbb{R})$. Generators denoted by $t$ with various indices
are nilpotent while $\{H^\prime,\,H_\Lambda\}$ are the diagonal
(non--compact) Cartan generators. The part of the algebraic
structure of $Solv_7$ which is not straightforward to deduce
consists in the non--vanishing commutation relations between the
generators
$\{t^{\Lambda\Sigma\tau},\,t_{\Lambda\Sigma},\,t^\tau\}$:
\begin{eqnarray}\label{E7-positive} {[}t^{\Lambda\Sigma
\tau},\,t^{\Gamma\Omega \sigma}]&=& \frac{1}{2}
\,\varepsilon^{\tau\sigma}\,
\varepsilon^{\Lambda\Sigma\Gamma\Omega\Pi\Delta}
\,t_{\Pi\Delta}\;,
\nonumber \\
{[}t_{\Lambda\Sigma},\,t^{\Gamma\Omega \tau}]&=& 2\,
\delta_{\Lambda\Sigma}^{\Gamma\Omega} \,t^\tau\;. \end{eqnarray}
 In order to define the involutions which give rise, upon truncation, to the two $N=4$
 models discussed above, let
us list in  Table \ref{table1} the bosonic q--form fields  of the
$N=8$ model with the gradings with respect to $(-1)^{H+q}$ and
$(-1)^{H^\prime+q}$ where the two Cartan generators $H$ and
$H^\prime$ are defined by the property that $C_{\mu\nu}^*$ be the
scalar with the highest $H$--grading equal to 2 and $C_{(0)}$ be
the scalar with the highest $H^\prime$--grading equal to 2 (in
other words they are the Cartan generators corresponding to the
positive roots which define the shift generators of $C_{\mu\nu}^*$
and $C_{(0)}$).
\begin{table}
\vskip -25pt
\begin{center}
\begin{tabular}{|c|c|c|}\hline
  field & $(-1)^{H+q}$& $(-1)^{H^\prime+q}$ \\\hline
  $\phi $ & + & + \\
  $g_{\Lambda\Sigma} $ & + & + \\
  $C_{(0)} $ & - & + \\
 $B_{\Lambda\Sigma} $ & - & - \\
  $C_{\Lambda\Sigma} $ & + & - \\
  $C_{\Lambda\Sigma\Gamma\Delta} $ & - & + \\
  $B_{\mu\nu}^* $ & - & -\\
  $C_{\mu\nu}^* $ & + & - \\
  $G^{\Lambda}_\mu $ & + & - \\
  $B_{\Lambda\mu} $ & - & + \\
  $C_{\Lambda\mu} $ & + & + \\
  $C_{\Lambda\Sigma\Gamma\mu} $ & - & - \\ \hline
\end{tabular}
\end{center}\caption{\label{table1} two orientifold involutions on the $N=8$ bosonic sector.}
\end{table}
 Recall that in our conventions the Kaluza--Klein vectors are denoted by
  $G^\Lambda_\mu=g^{\Lambda\Sigma}\,g_{\Sigma\mu}$. If we restrict
to fields with positive grading with respect to $(-1)^{H+q}$ we
obtain the bosonic part of the bulk sector of Type I theory on
$T^6$ while restricting to those with positive grading with
respect to $(-1)^{H^\prime+q}$ the resulting fields fit the bulk
sector of Type IIB theory on $T^6/\mathbb{Z}_2$ orientifold. In
\cite{dwst} the gauged $N=8$ supergravity describing the
low--energy limit of Type IIB on $T^6$ with fluxes was constructed
and it was shown how the various $N=4$
 gauged models constructed in \cite{aft1,aft2} and describing Type IIB on
 $T^{p-3}\times
T^{9-p}/\mathbb{Z}_2$ orientifolds in the presence of $Dp$--branes
and fluxes, could be obtained as consistent truncations of the
former model (the $p=3$ and $p=9$ cases correspond to the two
$N=4$ models that we are discussing in some detail in this
section). These truncations in the various cases can be
associated, through the recipe discussed above, to the Cartan
generator $H_{p-3}$ corresponding to the positive root which
defines the axion $C_{(p-3)}$, with components along the Neumann
directions $T^{p-3}$. With respect $(-1)^{H_{p-3}}$, for instance,
 a R--R $q+1$--form field strength and the Kalb--Ramond field strengths have grading:
\begin{eqnarray}
F_{\mu_1\dots\mu_{q+1}\,i_1\dots i_{k_+}\,a_1\dots
a_{k_-}}&\leftrightarrow&(-1)^{H_{p-3}}=(-1)^{\frac{1}{2}\,(7-p+k_+-k_--q)}\,,\nonumber\\
{\Scr H}_{\mu_1\dots\mu_{q+1}\,i_1\dots i_{k_+}\,a_1\dots
a_{k_-}}&\leftrightarrow&(-1)^{H_{p-3}}=(-1)^{\frac{1}{2}\,(k_+-k_--q)}\,;\,\,k_++k_-+q=2\,,
\end{eqnarray}
where $k_+,\,k_-$ are the number of indices along the Neumann and
Dirichlet directions of the torus respectively. As for the metric
and Kaluza--Klein vectors, $g_{ij},\, g_{ab}$ have
$(-1)^{H_{p-3}}$--grading $+$, $g_{ia}$ have grading -, $G^i_\mu$
grading - and $G^a_\mu$ grading +.
\par Below we give, for completeness, the kinetic matrix ${\Scr
N}$ of the vector fields in the two $N=4$ models, including the
respective boundary degrees of freedom. The kinetic terms for the
vector fields have the general form:
\begin{equation}
{\rm Im} \, {\Scr N}_{NM} \, F_{\mu\nu}^N F^{M \, \mu\nu} +
{\frac{1}{2}}\, {\rm Re} \, {\Scr N}_{NM} \,
\epsilon^{\mu\nu\rho\sigma} F^N_{\mu\nu} F^M_{\rho\sigma} \,.
\end{equation}
where $N,M$ run over the total number of vector fields. Let us
consider first the case of Type IIB on $T^6$ orientifold in the
presence of $D9$--branes. We denote by ${\Scr F }^i,\,F_i,\,F^v$
the field strengths of $G^i_\mu,\,C_{i\mu},\,A^v_\mu$ respectively
(where $v$ in this section only runs over the number $n_9$ of
$D9$--branes, $A^v_\mu$ and $a^v_i$ are the space--time and $T^6$
components of the gauge vectors on the boundary theory
respectively). From the index structure of the electric--magnetic
field strengths $({\Scr F }^i,\,F_i,\,F^v,\tilde{{\Scr F
}}_i,\,\tilde{F}^i,\,\tilde{F}_v)$ we readily see that the whole
group $\SO(6,6+n_9)$ is a global symmetry of the Lagrangian since
it  has a duality action which is block diagonal, namely has a
separate linear action on the electric and magnetic components. On
the other hand the $\SL(2,\mathbb{R})$ factor in the duality
group, which acts transitively on the scalars
$\{V_{6}\,e^{\frac{\phi}{2}},\, C_{\mu\nu}^*\}$ ($V_{6}$ being the
volume of $T^6$ ), is non perturbative since $({\Scr F
}^i,\,\tilde{F}^i)\in {\bf (2,6^\prime )}_{+1}$ and
$(F_i,\,\tilde{{\Scr F }}_i)\in {\bf (2,6 )}_{-1}$ with respect to
$\SL(2,\mathbb{R})\times \GL(6,\mathbb{R})$. This group therefore
is not a symmetry of the Lagrangian. The matrix ${\Scr N}_{MN}$
is:
\begin{eqnarray}
{\Scr N}&=&\,-c\,\eta-{\rm i}\, e^{\phi}\,L^{-1T}\,
L^{-1}\,=\,\left(\matrix{{\Scr N}_{ij}& {\Scr N}_i{}^j& {\Scr
N}_i{}^v\cr * & {\Scr N}^{ij}& {\Scr N}^{i v}\cr * & * & {\Scr
N}^{uv}}\right)\nonumber\\
{\Scr N}_{ij}&=&-{\rm
i}\,e^{\phi}\,(E_i^{\hat{k}}\,E_j^{\hat{k}}+\tilde{C}_{ik}\,\tilde{C}_{j\ell}\,
E^{-1}{}^k{}_{\hat{k}}\,E^{-1}{}^\ell{}_{\hat{k}} +a_i^v\,a_j^v)\nonumber\\
{\Scr N}_i{}^j&=&-c\, \delta_i^j-{\rm i}\,e^{\phi}\,\tilde{C}_{i k
}\, E^{-1}{}^k{}_{\hat{k}}\,E^{-1}{}^j{}_{\hat{k}}\nonumber\\
{\Scr N}_i{}^v&=& {\rm i}\,e^{\phi}\,(\tilde{C}_{i k }\,
E^{-1}{}^k{}_{\hat{k}}\,E^{-1}{}^j{}_{\hat{k}}\, a_{j}^v+a_i^v)\nonumber\\
{\Scr N}^{ij}&=& -{\rm
i}\,e^{\phi}\,E^{-1}{}^i{}_{\hat{k}}\,E^{-1}{}^j{}_{\hat{k}}\nonumber\\
{\Scr N}^{i v} &=& {\rm
i}\,e^{\phi}\,E^{-1}{}^i{}_{\hat{k}}\,E^{-1}{}^j{}_{\hat{k}}\, a_j^v\nonumber\\
{\Scr N}^{uv}&=& c\,\delta^{uv}-{\rm i}\, e^{\phi}\,(\delta^{uv}+a_i^u\,a_j^v\,E^{-1}{}^i{}_{\hat{k}}\,E^{-1}{}^j{}_{\hat{k}})\nonumber\\
\tilde{C}_{ij}&=&C_{ij}+\frac{1}{2}\,a_i^v\,a_j^v\,\,\,;\,\,\,\,\,c\equiv
C_{\mu\nu}^*\,\,;\,\,\, \eta_{MN}=\left(\matrix{0 &
\bfone_{6\times 6} & 0\cr \bfone_{6\times 6} & 0 & 0\cr 0 & 0&
-\bfone_{n_9\times n_9} }\right)
\end{eqnarray}
where $L$ denotes the coset representative of
$\SO(6,6+n_9)/[\SO(6)\times \SO(6+n_9)]$ and $E$ the coset
representative of $\GL(6,\mathbb{R})/\SO(6)$.\par Let us now
consider the $T^6/\mathbb{Z}_2$ orientifold model in the presence
of $n_3$ $D3$--branes and denote by ${F}_{a\tau},F^v$ the field
strengths of $B_{\mu a \tau},\,A^r_\mu$ respectively (where $r$
runs over the number $n_3$ of $D3$--branes, $A^r_\mu$ are the
vectors on the boundary theory and $a^{ar}$ the $T^6$ coordinates
of the $r^{th}$ $D3$--brane). As it is apparent from the index
structure of the electric--magnetic field strengths with respect
to $\SL(2,\mathbb{R})\times \GL(6,\mathbb{R})\times \SO(n_3)$ only
the subalgebra $gl(6,\mathbb{R})+ so(n_3)+{\bf
(15^\prime,1)}_{+2}+{\bf (6^\prime,1)}_{+1}$ generates the global
symmetry group of the Lagrangian. The components of the matrix
${\Scr N}$ are:
\begin{eqnarray}
{\Scr N}^{(1a)(1b)}&=&-i e^{-\phi}\,(E^{-1}\, E^{-1
T})^{ab}\nonumber\\
{\Scr N}^{(2a)(1b)}&=&i C_{(0)}\, e^{-\phi}\,(E^{-1}\, E^{-1
T})^{ab}-\tilde{C}^{ab}\nonumber\\
{\Scr N}^{(2a)(2b)}&=&-i \left[e^{\phi}\,(E^{-1}\, E^{-1 T}+a
a^T)^{ab}+e^{-\phi}\,C_{(0)}^2\,(E^{-1}\, E^{-1
T})^{ab}\right]-C_{(0)}\,(aa^T)^{ab}\nonumber\\
{\Scr N}^{(1a)\,r}&=&-a^{ar}\nonumber\\
{\Scr N}^{(2a)\,r}&=&a^{ar}\,(C_{(0)}+i e^{\phi})\nonumber\\
{\Scr N}^{rs}&=&-\delta^{rs}(C_{(0)}+i e^{\phi})\nonumber\\
\tilde{C}^{ab}&=& B^{ab}-\frac{1}{2}\,(aa^T)^{ab}
\end{eqnarray}
where $B^{ab}$ are related to $C_{abcd}$ by duality on the
internal torus.
\section{Alekseevski structure of a homogeneous quaternionic
algebra}\label{ale} In this section we shall briefly review the
structure of the solvable Lie algebra generating certain
homogeneous quaternionic manifolds ${\cal M}_Q$
(\emph{quaternionic algebra}) \cite{alek,c,dwvp1,dwvp2}. This will
be useful in order
\begin{itemize}
\item[i)]{ to define, using involutions of the algebra,
truncations of the manifold corresponding for instance to the
$N=2$ ``c--dual'' Special K\"ahler manifold or to the $N=1$
K\"ahler manifold arising from Calabi--Yau orientifold
compactifications \cite{bghl,bbhl,ggjl,dft} of Type IIB theory;}
\item[ii)]{to study the relations between the geometrical
properties of these truncations and in particular to show that for
symmetric manifolds  the $N=2$ Special K\"ahler  and the dual
$N=1$ K\"ahler manifolds are of the same kind.}
\end{itemize}
We shall restrict ourselves to ``Very Special'' homogeneous
quaternionic manifolds of rank 4. Let the quaternionic dimension
of ${\cal M}_Q$ be $n+1$.
 The corresponding quaternionic algebra $V$ has the form:
 \begin{eqnarray}
{\cal M}_Q&=&\exp(V)\,;\,\, V=U+\tilde{U}\nonumber\\
\left[U,U\right]&=&
U\,\,;\,\,\,\left[U,\tilde{U}\right]=\tilde{U}\,\,;\,\,\,\left[\tilde{U},\tilde{U}\right]=U
 \end{eqnarray}
 where $U$ is an algebra generating a K\"ahler submanifold and
 is stable with respect to the action of a complex structure
 $J_1$: $J_1\, U=U$. $\tilde{U}$ is related to $U$ by the action
 of a second complex structure $J_2$: $\tilde{U}=J_2\, U$, which,
 together with $J_1$ and $J_3=J_1 J_2$ generates the quaternionic
 structure of ${\cal M}_Q$.
$U$ has a linear adjoint action on the space $\tilde{U}$ which is
symplectic with respect to a suitable form $\hat{J}$ expressed in
terms of $J_1$. The structure of $U$ can be represented as
follows:
\begin{eqnarray}
U&=& F_0+F_1+F_2+F_3+X^\pm+Y^\pm+Z^\pm\nonumber\\
\left[F_I,\,F_J\right]&=&0\,\,\,\,\,\,\,\,I,J=0,1,2,3\nonumber\\
F_I&=&\{h_I,\,g_I\}\,\,;\,\,\,\,\left[h_I,\,g_I\right]=g_I
\end{eqnarray}
the solvable subalgebras $F_I$ generate four
$\SL(2,\mathbb{R})/\SO(2)$ submanifolds of ${\cal M}_Q$. The
generators $h_I$ define the Cartan subalgebra of $V$. The six
spaces $X^\pm,\,Y^\pm,\,Z^\pm$ consist of nilpotent  generators
and ${\rm dim} X^\pm+{\rm dim} Y^\pm+{\rm dim} Z^\pm=n-3 $.\par We
shall in general denote by $W(Solv)$ the normal K\"ahler manifold
generated by the solvable Lie algebra $Solv$. In particular
$W(F_1+F_2+F_3+X^\pm+Y^\pm+Z^\pm)$ is the Special K\"ahler
manifold which corresponds to ${\cal M}_Q$ through the c--map.\par
The generators of $\tilde{U}$, denoted by
$\{p^I,\,q_I,\,\tilde{X}^\pm,\,\tilde{Y}^\pm,\,\tilde{Z}^\pm\}$
can be arranged in a symplectic vector with respect to the adjoint
action of $U$:
\begin{eqnarray}
\left(\matrix{v^\lambda\cr
w_\sigma}\right)\,\,;\,\,\,v^\lambda=\left(\matrix{p_0\cr
p_\alpha\cr \tilde{X}^-\cr\tilde{Y}^- \cr
\tilde{Z}^-}\right)\,\,;\,\,\,w_\sigma=\left(\matrix{q_0\cr
q_\alpha\cr \tilde{X}^+\cr\tilde{Y}^+ \cr \tilde{Z}^+}\right)
\end{eqnarray}
The algebraic structure of the whole algebra $V$ can be deduced
from the gradings of the various nilpotent generators with respect
to $h_I$ which are listed in Table \ref{table2}.
\begin{table}
\begin{center}
\begin{tabular}{|c|c|c|c|c|}\hline
gen.& field & grading & H-grading & ${\rm
H}^\prime$--grading\\\hline
  $g_0$ & $B^*_{\mu\nu}$ & $(1,0,0,0)$ & 1 & 1  \\
  $g_1$ & $B_{(2)}$ & $(0,1,0,0)$ & 1 & -1 \\
  $g_2$ & $B_{(2)}$ & $(0,0,1,0)$ & 1 & -1  \\
  $g_3$ & $B_{(2)}$ & $(0,0,0,1)$ & 1 & -1  \\
 $X^\pm$ & $\cases{B_{(2)}\cr g}$ & $(0,0,\frac{1}{2},\pm \frac{1}{2})$ &
 $\cases{1\cr 0}$ & $\cases{-1\cr 0}$ \\
  $Y^\pm$ & $\cases{B_{(2)}\cr g}$ & $(0,\frac{1}{2},0,\pm \frac{1}{2})$ &
   $\cases{1\cr 0}$ & $\cases{-1\cr 0}$ \\
  $Z^\pm$ & $\cases{B_{(2)}\cr g}$ & $(0,\frac{1}{2},\pm \frac{1}{2},0)$ &
   $\cases{1\cr 0}$ & $\cases{-1\cr 0}$\\
  $p_0$ & $C^*_{\mu\nu}$ & $(\frac{1}{2},\frac{1}{2},\frac{1}{2},\frac{1}{2})$ & 2 & -1 \\
 $q_0$ & $C_{(0)}$ & $(\frac{1}{2},-\frac{1}{2},-\frac{1}{2},-\frac{1}{2})$ & -1 & 2  \\
 $p_1$ & $C_{(2)}$ & $(\frac{1}{2},\frac{1}{2},-\frac{1}{2},-\frac{1}{2})$ & 0 & 1  \\
  $q_1$ & $C_{(4)}$ & $(\frac{1}{2},-\frac{1}{2},\frac{1}{2},\frac{1}{2})$ & 1 & 0 \\
 $p_2$ & $C_{(2)}$ & $(\frac{1}{2},-\frac{1}{2},\frac{1}{2},-\frac{1}{2})$ & 0 & 1  \\
  $q_2$ & $C_{(4)}$ & $(\frac{1}{2},\frac{1}{2},-\frac{1}{2},\frac{1}{2})$ & 1 & 0 \\
   $p_3$ & $C_{(2)}$ & $(\frac{1}{2},-\frac{1}{2},-\frac{1}{2},\frac{1}{2})$ & 0 & 1 \\
   $q_3$ & $C_{(4)}$ & $(\frac{1}{2},\frac{1}{2},\frac{1}{2},-\frac{1}{2})$ & 1 & 0  \\
 $\tilde{X}^\pm$ & $\cases{C_{(4)}\cr C_{(2)}}$ & $(\frac{1}{2},\pm \frac{1}{2},0,0)$ &
 $\cases{1\cr 0}$ & $\cases{0\cr 1}$ \\
  $\tilde{Y}^\pm$ & $\cases{C_{(4)}\cr C_{(2)}}$ & $(\frac{1}{2},0,\pm \frac{1}{2},0)$ &
 $\cases{1\cr 0}$ & $\cases{0\cr 1}$ \\
 $\tilde{Z}^\pm$ & $\cases{ C_{(4)}\cr C_{(2)}}$ & $(\frac{1}{2},0,0,\pm \frac{1}{2})$ &
 $\cases{1\cr 0}$ & $\cases{0\cr 1}$\\\hline
\end{tabular}\caption{\label{table2}generators of $V$ and their gradings.}
\end{center}\end{table}
In this list we have also specified for each nilpotent generator
the scalar field which parametrizes it in terms of the
corresponding ten--dimensional parent fields: $g$ represents the
metric, $B_{(2)},\,C_{(2)},\,C_{(4)}$ the Type IIB forms. This
generator--field correspondence  can be justified as follows. We
start from knowing that $U$ is parametrized   by NS--NS fields
coming either from $B_{(2)}$ or from the metric moduli $g$ and the
dilaton $\phi$. The generators related to $B_{(2)}$ ($T^B$) can be
identified form the typical commutation property with those
corresponding to metric moduli ($T^g$) or dilaton ($T^\phi=h_0$)
\begin{eqnarray}
\left[T^B,T^g\right]&=&T^B\,\,;,\,\,\,\left[T^B,T^\phi\right]=T^B
\end{eqnarray}
Thus we have that
\begin{eqnarray}
\{T^B\}&=&\{g_0,g_\alpha,X^+, \,Y^+,\,Z^+\}\nonumber\\
\{T^\phi,T^g\}&=&\{h_0,\,h_\alpha,\,X^-, \,Y^-,\,Z^-\}
\end{eqnarray}
 If we denote by the generic
symbols $T^{C_0},T^{C_2},T^{C_2^*},T^{C_4}$ the nilpotent
generators parametrizing the R--R scalars coming from $C_{(0)},
C_{(2)}, C_{\mu\nu}^*,C_{(4)}$, the following general commutation
properties can help us defining the (qualitative) correspondence:
\begin{eqnarray}
\left[T^B,T^{C_0}\right]&=&T^{C_2}\,\,;\,\,\,
\left[T^B,T^{C_2}\right]=T^{C_4}\,\,;\,\,\,\left[T^B,T^{C_4}\right]=T^{C_2^*}\nonumber\\
\left[T^{C_2},T^{C_4}\right]&=&\{g_0\}\,\,;\,\,\,
\left[T^{C_0},T^{C_2^*}\right]=\{g_0\}\,\,;\,\,\,
\end{eqnarray}
To make contact with the notation used in the literature (see
\cite{dwvp1,dwvp2,dft}) for the coordinates and solvable
isometries of the quaternionic manifold let us specify their
correspondence with the solvable generators used here:
\begin{eqnarray}
\{h_0,g_0,q_0,p_0\}&\leftrightarrow
&\{D,\tilde{\Phi},\zeta^0,\tilde{\zeta}_0\}\nonumber\\
\{q_\alpha,\tilde{X}^+,\tilde{Y}^+,\tilde{Z}^+\}&\leftrightarrow
&\{\tilde{\zeta}_a\}\nonumber\\
\{p_\alpha,\tilde{X}^-,\tilde{Y}^-,\tilde{Z}^-\}&\leftrightarrow
&\{\zeta^a\}\nonumber\\
\{h_\alpha,X^-,Y^-,Z^-\}&\leftrightarrow
&\{y^a\}\nonumber\\
\{g_\alpha,X^+,Y^+,Z^+\}&\leftrightarrow &\{x^a\}\nonumber\\
&&\nonumber\\
\{q_\alpha,\,\tilde{X}^+,\tilde{Y}^+,\tilde{Z}^+\}&=&\{\beta_a\}\,;\,\,\,
\{p_\alpha,\,\tilde{X}^-,\tilde{Y}^-,\tilde{Z}^-\}=\{\alpha^a\}\nonumber\\
q_0&=&\alpha^0\,;\,\,\,p_0=\beta_0
\end{eqnarray}
 As far as the $X,\,Y,\,Z$ generators are concerned
the quaternionic structure of the manifold requires that either
${\rm dim}(X^\pm)=0\,\mbox{mod}\, 4$ or if ${\rm dim}(X^\pm)\neq
0\,\mbox{mod}\, 4$ then we should have ${\rm dim}(Y^\pm)={\rm
dim}(Z^\pm)$ \cite{dwvp1,dwvp2}. From the grading structure we
also deduce that $\left[X^-,Z^-\right]=Y^-$.
\par As far as the ${\rm dim}(X^\pm)= 0$ case is concerned, if
$p={\rm dim}(Y^\pm)$ and $q={\rm dim}(Z^\pm)$ the K\"ahler algebra
is denoted by $K(p,q)$ and the quaternionic algebra by $W(p,q)$.
\par In the $D3/D7$ problem the scalar manifold of the vector
multiplet sector is generated by an algebra $K(p,q)$, where
$p=n_3$ and $q=n_7$. We shall come back to these manifolds in the
next section.
\paragraph{Scaling symmetries and truncations}
 Let us consider three relevant truncations of the scalar sector
 defined by three different Cartan generators, according to the
 recipe given in section \ref{truncation}.\par
If we consider $h=2 h_0$ we see that all the R--R generators are
odd  and the truncation leaves the generators $h_0, g_0$ and the
K\"ahler algebra obtained under c-map:
\begin{eqnarray}
W(h_0,,g_0)&\times &W\nonumber\\
W&=&W(g_\alpha,\,h_\alpha,\,X^\pm, \,Y^\pm,\,Z^\pm)
\end{eqnarray}
If we consider $h=H=h_0+h_1+h_2+h_3$, from the table above we
deduce that the truncation leaves:
\begin{eqnarray}
W(h^\prime_0,p_0)&\times &W_1 \nonumber\\
W_1&= & W(h^\prime_\alpha,\,X^-,\,Y^-,\,Z^-, p_\alpha,\tilde{X}^-,
\,\tilde{Y}^-,\,\tilde{Z}^-)
\end{eqnarray}
corresponding to the fields:
\begin{eqnarray}
\{\phi,G,C^*_{\mu\nu},C_{(2)}\}
\end{eqnarray}
where $h^\prime_I$ are suitable linear combinations of $h_I$. The
manifold $W_1$ is still Special K\"ahler and defines the vector
multiplet sector of the sigma model obtained by reducing Type I
theory on a Calabi--Yau manifold.
\par For each nilpotent isometry generator $A$ we may formally
consider its \emph{negative} $\hat{A}$ defined by opposite
$h_I$--grading (root) with respect to $A$ \footnote{Notice that in
the literature (see for instance \cite{dwvp1,dwvp2,dft}) the
hatted symbols are used to denote generators which have opposite
$h_0$ grading but same $h_\alpha$ grading with respect to the
\emph{positive} counterparts (in the non--homogeneous cases
$h_\alpha$ do not always exist). Here for convenience we used this
notation for generators with opposite $h_0$ and
$h_\alpha$--grading. Therefore the nilpotent generators of
$\SL(2,\mathbb{R})_{{\rm IIB}}$ are in the two notations:
$q_0=\alpha^0$, $\hat{q}_0=\hat{\beta}_0$.}. For non--symmetric
manifolds $\hat{A}$ is in general not an isometry. If $\hat{A}$ is
an isometry, then we can define the Weyl transformation ${\cal
O}_A=\exp(\frac{\pi}{2} (A-\hat{A}))$ which, besides being an {\it
automorphism} of the isometry algebra, namely preserving the whole
algebraic structure of $V$, maps Cartan generators into Cartan
generators. For very special manifolds the generator $\hat{q}_0$
is always an isometry since $h_0, q_0, \hat{q}_0$ generate the
${\rm SL}(2,\mathbb{R})_{{\rm IIB}}$ ten--dimensional type IIB
duality and the mapping between $W$ and $W_1$ is provided by the
discrete Weyl transformation ${\cal O}_{q_0}$:
\begin{eqnarray}
{\cal O}_{q_0}\,W(h_0,g_0)\,{\cal O}_{q_0}^{-1}&=&W(h^\prime_0,p_0)\nonumber\\
{\cal O}_{q_0}\,W\,{\cal O}_{q_0}^{-1}&=&W_1
\end{eqnarray}
Since the two algebras are mapped into each other by an
automorphism of the quaternionic algebra which is also an
isometry, and since the geometric properties of these manifolds
are encoded the commutation relations between their generators and
the remaining generators in $V$, we deduce that the $W_1$ should
be a special K\"ahler manifold as well, of the same kind as
$W$.\par
 If we consider $h=H^\prime=h_0-h_1-h_2-h_3$, from Table
 \ref{table2} we deduce that the truncation leaves:
\begin{eqnarray}
W(h^{\prime\prime}_0,q_0)&\times &W_2\nonumber\\ W_2 & = &
W(h^{\prime\prime}_\alpha,\,X^-, \,Y^-,\,Z^-,
q_\alpha,\tilde{X}^+, \,\tilde{Y}^+,\,\tilde{Z}^+)
\end{eqnarray}
corresponding to the fields:
\begin{eqnarray}
\{\phi,G,C_{(0)},C_{(4)}\}
\end{eqnarray}
The generators $h^{\prime\prime}_I$ are related to $h_I$ by a
linear transformation. The manifold $W_2$ defines the sigma model
of the $N=1$ theory obtained by reduction of Type IIB supergravity
on a Calabi--Yau orientifold. It was constructed in
\cite{bbhl,ggjl} ad it is referred to as "dual'' K\"ahler
manifold. The relation between $W$ and $W_2$ would involve the
action of generators corresponding to negative roots other than
$\hat{q}_0$ and which are present among the isometries of the
manifold only for symmetric manifolds. Let us show this. Suppose
first that the special K\"ahler manifold $W$ is symmetric. In this
case, for each nilpotent generator $A$ in its solvable algebra the
corresponding {\it negative} counterpart $\hat{A}$ is an isometry.
In particular the generators $\hat{g}_\alpha$ are isometries and
so are the Weyl transformations ${\cal O}_{g_\alpha}$. Let us
prove that the following Weyl transformation:
\begin{eqnarray}
{\cal O}&=&{\cal O}_{g_3}{\cal O}_{g_2}{\cal O}_{g_1}{\cal
O}_{q_0}
\end{eqnarray}
which is an isometry, maps the dual K\"ahler manifolds into each
other:
\begin{eqnarray}
{\cal O}\,W(h_0,g_0)\,{\cal O}^{-1}&=&W(h^{\prime\prime}_0,q_0)\nonumber\\
{\cal O}\,W\,{\cal O}^{-1}&=&W_2 \label{dualcor}
\end{eqnarray}
Indeed by inspection of the various gradings we can show that:
\begin{eqnarray}
\matrix{&g_0& g_1 & g_2 & g_3 & X^+ & Y^+& Z^+\cr {\cal O}_{q_0}
&\downarrow & \downarrow & \downarrow & \downarrow & \downarrow &
\downarrow & \downarrow\cr &p_0& p_1 & p_2 & p_3 & \tilde{X}^- &
\tilde{Y}^-& \tilde{Z}^-\cr {\cal O}_{g_1} &\downarrow &
\downarrow & \downarrow & \downarrow & \downarrow & \downarrow
&\downarrow\cr &q_1& q_0 & q_3 & q_2 &\tilde{X}^+ & \tilde{Y}^-&
\tilde{Z}^-\cr {\cal O}_{g_2} &\downarrow & \downarrow &
\downarrow & \downarrow & \downarrow & \downarrow & \downarrow\cr
&p_3& p_2 & p_1 & p_0 &\tilde{X}^+ & \tilde{Y}^+& \tilde{Z}^-\cr
{\cal O}_{g_3} &\downarrow & \downarrow & \downarrow & \downarrow
& \downarrow & \downarrow & \downarrow\cr &q_0& q_1 & q_2 & q_3
&\tilde{X}^+ & \tilde{Y}^+& \tilde{Z}^+}
\end{eqnarray}
The action of ${\cal O}$ on the other generators parametrized by
the dilaton and the metric moduli has the following effect:
\begin{eqnarray}
{\cal O}\,\{h_0,h_\alpha,X^-,Y^-,Z^-\}\,{\cal
O}^{-1}&=&\{h^{\prime\prime}_0,h^{\prime\prime}_\alpha,\hat{X}^-,\hat{Y}^-,\hat{Z}^-\}
\end{eqnarray}
due to the symmetry property of $W$ the generators
$\{h_0,h_\alpha,\hat{X}^-,\hat{Y}^-,\hat{Z}^-\}$ are still in the
isometry group of the dual K\"ahler manifold $W_2$. This proves
eqs. (\ref{dualcor}) and thus that:\par \emph{If the original
Special K\"ahler manifold $W$ is symmetric, the dual one $W_2$
will be a Special K\"ahler manifold of the same kind.} In Appendix
\ref{appendiceA} we shall derive the Riemann tensor for the dual
K\"ahler manifold and show that it coincides which the one of the
original Special K\"ahler manifold in the symmetric case.
\par
 Finally
it can be shown that if $W$ is symmetric the full quaternionic
algebra is symmetric as well. Indeed by successive action of
$\hat{g}_\alpha$ on $\hat{q}_0$ we generate $\hat{p}_0$ which is
thus also an isometry. The commutator
$\left[\hat{q}_0,\hat{p}_0\right]=\hat{g}_0$, denoted in the
literature  by $\epsilon^-$, and all the other {\it negative}
counterparts of the quaternionic generators can be generated as
commutators of isometries, thus belonging to the isometry algebra
as well.\par Manifolds which are ``dual'' but not symmetric are in
general different. As an example we can consider ``dual'' K\"ahler
manifolds which are homogeneous but not symmetric. For instance we
may choose $W=K(1,1)$ of complex dimension $5$. To show that the
two manifolds are different it suffices to show that the
corresponding scalar curvatures are different. Indeed one finds:
\begin{eqnarray}
R(W)&=&-18\,\,;\,\,\,\,R(W_2)=-16\,.
\end{eqnarray}
\section{Special coordinates, solvable coordinates and B.I. action}
\label{veryspecial} The prepotential for the spacial geometry of
the $D3-D7$ system is
\begin{eqnarray}
{\cal F}&=&stu-\frac{1}{2}\,s (x^i)^2-\frac{1}{2}\,u
(y^r)^2\,.\label{prepo}
\end{eqnarray}
This prepotential was obtained in \cite{adft}, by using arguments
based on duality symmetry, four dimensional Chern--Simons terms
coming from the p--brane couplings as well as couplings of vector
multiplets in $D=4$ and $D=8$.\par A similar result was advocated
in \cite{fms,abfpt} by performing first a $K3$ reduction to $D=6$
and then further compactifying the theory to $D=4$ on $T^2$.\par
The subtlety of this derivation is that the naive Born--Infeld
action derived for $D5$ and $D9$ branes in $D=6$ gives kinetic
terms for the scalar fields which, at the classical level, are
inconsistent with $N=2$ supersymmetry. This is a consequence of
the fact that anomalies are present in the theory, as in the
$D=10$ case. The mixed anomaly local counterterms are advocated to
make the Lagrangian $N=2$ supersymmetric in $D=4$.\par Therefore
the corrected Lagrangian, in the original brane coordinates is
highly non--polinomial. In fact the original Born--Infeld,
Chern--Simons naive (additive) classical scalar action
\begin{eqnarray}
&&\frac{|\partial s^\prime+ c^r \partial
d^r|^2}{(s^\prime-\bar{s}^\prime)^2}+\frac{|\partial u^\prime+ a^i
\partial b^i|^2}{(u^\prime-\bar{u}^\prime)^2}+\frac{|t^\prime\,\partial d^r+\partial
c^r|^2}{(s^\prime-\bar{s}^\prime)\,(t^\prime-\bar{t}^\prime)}+\frac{|t^\prime\,\partial
b^i+\partial
a^i|^2}{(u^\prime-\bar{u}^\prime)\,(t^\prime-\bar{t}^\prime)}+\frac{|\partial
t^\prime|^2}{(t^\prime-\bar{t^\prime})^2}\nonumber\\
&&\phantom{aaaaa}s^\prime=s-\frac{1}{2}\,d^r
y^r\,\,;\,\,\,u^\prime=u-\frac{1}{2}\,b^i x^i\,\,;\,\,t^\prime =t\nonumber\\
&&\phantom{aaaaa}x^i=a^i+t \,b^i\,\,;\,\,\,y^r=c^r+t\,d^r\,,
\label{bia}
\end{eqnarray}
has a metric which was shown \cite{abfpt} to be K\"ahler with
K\"ahler potential\footnote{$Y_{SK}$ differs by a factor $-i$ from
the Special geometry formula obtained from the prepotential in
\ref{prepo}.}
\begin{eqnarray}
K&=&
-\log\left[(s-\bar{s})(t-\bar{t})-\frac{1}{2}\,(y^r-\bar{y}^r)^2\right]-
\log\left[(u-\bar{u})(t-\bar{t})-\frac{1}{2}\,(x^i-\bar{x}^i)^2\right]+\log (t-\bar{t})\nonumber\\
&&=-\log Y_{SK}-\log(1+\frac{X_4}{Y_{SG}})
\end{eqnarray}
where
\begin{eqnarray}
X_4&=&\frac{(x^i-\bar{x}^i)^2 (y^r-\bar{y}^r)^2}{4\,
(t-\bar{t})}\nonumber\\Y_{SK}&=&(s-\bar{s})(t-\bar{t})(u-\bar{u})-\frac{1}{2}\,(u-\bar{u})(y^r-\bar{y}^r)^2-\frac{1}{2}\,(s-\bar{s})(x^i-\bar{x}^i)^2\nonumber\\
\end{eqnarray}
where here and in the following summation over repeated indices is
understood.  Therefore the correction to the scalar metric in the
brane coordinates is:
\begin{eqnarray}
\partial_p\partial_{\bar{q}}\Delta K&=&\partial_p\partial_{\bar{q}}\log(1+\frac{X_4}{Y_{SG}})
\end{eqnarray}
 It is clear that the classical
brane coordinates are not good ``supersymmetric'' coordinates, in
that the corrected action is not polynomial in them. From the fact
that the combined system is a homogeneous space, we indeed expect
that suitable  coordinates exist such that the quantum corrected
($N=2$ supersymmetric) action has a simple polynomial dependence
on them, including the interference term. Such coordinates do
indeed exist and allow to write the combined Born--Infeld action
and supersymmetric counterterms, in a manifest supersymmetric way.
Modulo field redefinitions, these coordinates reduce to the
standard brane coordinates when either the $D3$ or the
$D7$--branes are absent, in which cases the homogeneous space
becomes a symmetric space. This parametrization in terms of
``supersymmetric'' coordinates, corresponds to the solvable Lie
algebra description of the manifold which we shall discuss in what
follows. In the previously introduced Alekseevski's notation the
manifold under consideration is of type $K(n_3,n_7)$ which can be
written as:
\begin{eqnarray}
K(n_3,n_7)&=& W(g_\alpha,\,h_\alpha,\,Y^\pm,\,Z^\pm)\nonumber\\
{\rm dim}(Y^\pm)&=&n_3\,\,;\,\,\,{\rm dim}(Z^\pm) =  n_7
\end{eqnarray}
where $n_3$ and $n_7$ denote the number of $D3$ and $D7$--branes
respectively. For the sake of clarity we rename in this section
the generators $h_1,h_2,h_3$ and $g_1,g_2,g_3$ by $h_t,h_u,h_s$
and $g_t,g_u,g_s$ respectively. Our identification of the scalar
fields with solvable parameters is described by the following
expression for a generic solvable Lie algebra element:
\begin{eqnarray}
Solv &=&\{\sum_{\alpha=t,u,s}\varphi^\alpha h_\alpha
+\hat{\theta}_t g_t+\theta_u g_u+\theta_s g_s+y^{r\pm
}Y^{\pm}_r+z^{i\pm
}Z^{\pm}_i\}\nonumber\\
&&\phantom{aaaaaa}\hat{\theta}_t = \theta_t+y^{r+}\,
y^{r-}+z^{i+}\, z^{i-}
\end{eqnarray}
where $(y^{r+},y^{r-})$ and $(z^{i+},z^{i-})$ are related to the
real and imaginary parts of the $D3$ and $D7$--branes complex
coordinates along $T^2$. The non trivial commutation relations
between the above solvable generators are:
\begin{eqnarray}
[h_t,Y^\pm]&=&
\frac{1}{2}\,Y^\pm\,\,\,;\,\,\,\,\,[h_t,Z^\pm]=\frac{1}{2}\,
Z^\pm\nonumber\\
\left[ h_s,Y^\pm \right] &=&\pm\frac{1}{2}\,Y^\pm\,\,\,;\,
\,\,\,\,\left[h_u,Z^\pm\right]=\pm\frac{1}{2}\,
Z^\pm\nonumber\\
\left[g_s,Y^-\right]&=&Y^+\,\,\,;\,\,\,\,\,\left[g_u,Z^-\right]=
Z^+\nonumber\\
\left[Y^+_r,Y^-_s\right]&=&\delta_{rs}\,g_t\,\,\,;\,\,\,\,\,\left[Z^+_i,Z^-_j\right]=
\delta_{ij}\,
g_t\,\,;\,\,\,r,s=1,\dots,n_3\,\,i,j=1,\dots,n_7\nonumber\\
\left[h_\alpha, g_\alpha\right]&=&g_\alpha\,\,;\,\,\,\alpha=t,u,s
\label{comkal}
\end{eqnarray}
We exponentiate the solvable algebra using the following
coset-representative:
\begin{eqnarray}
L&=&e^{\theta_s g_s}\,e^{y^{r-} Y^-_r}\,e^{y^{r+}
Y^+_r}\,e^{\theta_u g_u}\,e^{z^{i-} Z^-_i}\,e^{z^{i+} Z^+_i}\, e^{
\hat{\theta}_t\,g_t}\,e^{\varphi^\alpha h_\alpha}
\end{eqnarray}
The order of the exponentials in the coset representative and the
particular parameter $\hat{\theta}_t$ used for $g_t$, have been
chosen in such a way that the axions $\theta_s,
\,\theta_t,\,\theta_u,\,y^{r+},\,z^{i+}$ appear in the resulting
metric only covered by derivatives. The metric reads:
\begin{eqnarray}
ds^2&=&
(d\varphi_\alpha)^2+e^{-2\varphi_t}\,\left(d\theta_t+\frac{1}{2}\,d\theta_u
(z^{-})^2+\frac{1}{2}\,d\theta_s (y^{-})^2+z^{i-}\,
dz^{i+}+y^{r-}\, dy^{r+} \right)^2+\nonumber\\
&&
e^{-2\varphi_u}\,d\theta_u^2+e^{-2\varphi_s}\,d\theta_s^2+e^{-\varphi_t-\varphi_u}\,
(dz^{i+}+d\theta_u\,z^{i-})^2+e^{-\varphi_t+\varphi_u}\,
(dz^{i-})^2+\nonumber\\&& e^{-\varphi_t-\varphi_s}\,
(dy^{r+}+d\theta_s\,y^{r-})^2+e^{-\varphi_t+\varphi_s}\,
(dy^{r-})^2 \nonumber\\&&(z^{+})^2 \equiv
\sum_{i=1}^{n_7}(z^{i+})^2\,;
\,\,(y^{+})^2\equiv\sum_{r=1}^{n_3}(y^{r+})^2
\end{eqnarray}
Identifying the axionic coordinates
$\theta_s,\,\theta_t,\,\theta_u,\,y^{r+},\,z^{i+}$ with the real
part of the special coordinates $s,\,t,\,u,\,y^r,\,x^i$, and
comparing the corresponding components of the metric one easily
obtains the following relations between the solvable coordinates
and the special coordinates\footnote{We notice that in the
previous paper \cite{adft} the imaginary parts of $u$ and $t$ were
chosen to be positive. This however is inconsistent with the
positivity domain of the vector kinetic terms which requires
$s,t,u$ to have negative imaginary parts. Indeed $\Im s$ and $\Im
u $ appear as coefficients in the kinetic terms of the $D7$ and
$D3$--brane vectors.}:
\begin{eqnarray}
s&=&\theta_s-\frac{i}{2}\,e^{\varphi_s}\,\,;\,\,\,u=\theta_u-\frac{i}{2}\,
e^{\varphi_u}\nonumber\\
t&=&\theta_t-\frac{i}{2}\,\left(e^{\varphi_t}+\frac{1}{2}
\,e^{\varphi_u}\,(z^{-})^2+\frac{1}{2}\,e^{\varphi_s}\,
(y^{-})^2\right)\nonumber\\
x^i&=&z^{i+}+\frac{i}{2}\, e^{\varphi_u}\,
z^{i-}\,\,;\,\,\,\,y^r=y^{r+}+\frac{i}{2}\,e^{\varphi_s}\, y^{r-}
\end{eqnarray}
\par Let us notice that the classical B--I+C--S action (\ref{bia}),
with no interference term in the $D3$ ($c,\,d$) and $D7$ ($a,\,b$)
brane coordinates is still described by a homogeneous manifold
spanned by the following  $2\,n_3+2\,n_7+6$ isometries:
\begin{eqnarray}
u&\rightarrow & e^{\lambda_u}\,u\,\,;\,\,\,\delta u=u_0+a_0^i
b^i\nonumber\\
s&\rightarrow & e^{\lambda_s}\,s\,\,;\,\,\,\delta s=s_0+c_0^r
d_r\nonumber\\
t&\rightarrow & e^{\lambda_t}\,t\,\,;\,\,\,\delta t=t_0\nonumber\\
c^r&\rightarrow &
e^{\frac{\lambda_s+\lambda_t}{2}}\,c^r\,\,;\,\,\,\delta c^r=t_0\,
d^r\nonumber\\
d^r&\rightarrow &
e^{\frac{\lambda_s-\lambda_t}{2}}\,d^r\,\,;\,\,\,\delta
d^r=d_0^r\nonumber\\
a^i&\rightarrow &
e^{\frac{\lambda_u+\lambda_t}{2}}\,a^i\,\,;\,\,\,\delta
a^i=a_0^i+t_0\,
b^i\nonumber\\
b^i&\rightarrow &
e^{\frac{\lambda_u-\lambda_t}{2}}\,b^i\,\,;\,\,\,\delta
b^i=b_0^i\,.
\end{eqnarray}
The underlying homogeneous space is generated by the following
rank 3 solvable Lie algebra
$\{T_a^i,\,T_b^i,\,T_c^r,\,T_d^r,\,h_s,\,h_t,\,h_u,\,g_s,\,g_t,\,g_u\}$
whose non trivial commutation relations are:
\begin{eqnarray}
\left[T_a^i,\,T_b^j\right]&=&\delta^{ij}\,g_u\,\,;\,\,\,\,
\left[T_c^r,\,T_d^s\right]=\delta^{rs}\,g_s\nonumber\\
\left[T_b^i,\,g_t\right]&=&T_a^i\,\,;\,\,\,\,\left[T_d^r,\,g_t\right]=T_c^r\nonumber\\
\left[h_\alpha,\,g_\alpha\right]&=&g_\alpha\,\,\,\,\,\alpha=s,t,u\nonumber\\
\left[h_s,\,T^r_d\right]&=&\frac{1}{2}\,T^r_d\,\,;\,\,\,\,
\left[h_s,\,T^r_c\right]=\frac{1}{2}\,T^r_c\nonumber\\
\left[h_u,\,T^i_b\right]&=&\frac{1}{2}\,T^i_b\,\,;\,\,\,\,
\left[h_u,\,T^i_a\right]=\frac{1}{2}\,T^i_a\nonumber\\
\left[h_t,\,T^r_d\right]&=&-\frac{1}{2}\,T^r_d\,\,;\,\,\,\,
\left[h_t,\,T^r_c\right]=\frac{1}{2}\,T^r_c\nonumber\\
\left[h_t,\,T^i_b\right]&=&-\frac{1}{2}\,T^i_b\,\,;\,\,\,\,
\left[h_t,\,T^i_a\right]=\frac{1}{2}\,T^i_a\,,
\end{eqnarray}
where the nilpotent generators have been labelled by the
corresponding axionic scalar fields. This space is not a subspace
of the original quanternionic space, but it becomes so if we set
either $a,b=0$ and exchange the role of $s$ and $t$ or if we set
$c,d=0$ and exchange the role of $u$ and $t$.\par The amazing
story is that the coordinates in $D=4$ corresponding to the
supersymmetric theory, deform this space into an other homogeneous
space generated by the isometries in (\ref{comkal}) which
corresponds to an $N=2$ Special Geometry.\par The relation between
the solvable Lie algebra generators
$\{T_a^i,\,T_b^i,\,T_c^r,\,T_d^r,\,h_s,\,h_t,\,h_u$
$,\,g_s,\,g_t,\,g_u\}$ corresponding to the classical coordinates
and the solvable generators
$\{Y^\pm,\,Z^\pm,\,h_\alpha,\,g_\alpha\}$ corresponding to the
``supersymmetric'' coordinates is the following:
\begin{eqnarray}
T_a^i&=&\hat{Z}^{i+}\,\,;\,\,\,T_b^i=\hat{Z}^{i-}\nonumber\\
T_c^r&=&\hat{Y}^{r+}\,\,;\,\,\,T_d^r=\hat{Y}^{r-}
\end{eqnarray}
where $\hat{Y}$ and $\hat{Z}$ are the generators with opposite
grading with respect to $Y$ and $Z$ respectively. It can be shown
that in the manifold $K(n_3,n_7)$,  $\hat{Y}$ or $\hat{Z}$ are
isometries only if $n_7=0$ or $n_3=0$ respectively. Indeed in
these two cases the manifold is symmetric and each solvable
nilpotent isometry has a ``hidden'' counterpart with opposite
grading. Otherwise the manifold spanned by the classical
coordinates and the manifold parametrized by the
``supersymmetric'' ones are in general different.
\section{Discussion on the $D=6$ dimensional origin of the
prepotential} The cubic prepotential for the $D3,\,D7$ branes can
be obtained in a number of ways by using different string
dualities.\par  Type I theory, obtained as a Type IIB orientifold
on $K3$, in the presence of $D5$ and $D9$ branes has an effective
description in terms of an $N=1$ theory in $D=6$ \cite{abfpt}. In
this theory there are $n_t$ tensor multiplets and $n_v$ vector
multiplets ($n_t=1$ in models equivalent to perturbative heterotic
string). Upon further compactification on $S^1$ to $D=5$, one
obtains a Very Special geometry of the type \cite{fms}
\begin{eqnarray}
{\cal F}&=&z\, b^r\eta_{rs} b^s-b^r\, C_{rxy} a^x
a^y\nonumber\\&&x,y=1,\dots, n_v\,;\,\,\,r,s=1,\dots , n_t+1\,,
\end{eqnarray}
where the physical scalar fields are $n_t+n_v+1$, of which $n_t$,
described by components of $b_r$, come from the tensor multiplets
in $D=6$, $n_v$, namely $a^x$, from the $D=6$ vectors and one
coincides with the Kaluza--Klein scalar $z$ ($\eta_{rs}$ is the
lorentzian metric with signature $(1,n_t)$ of the tensor
multiplets). For $n_t=1$
\begin{eqnarray}
{\cal F}&=&z\, b\,c-b\, v_{xy} a^x a^y-c\,\tilde{v}_{xy} a^x a^y
\end{eqnarray}
and we obtain the Very Special Geometry described in section
\ref{veryspecial}. This result for $n_t=1$ is ``dual'' to
heterotic string on $K3\times S^1$ or M--theory on a Calabi--Yau
threefold which is an elliptic fibration \cite{fms}. The case
$n_t>1$ corresponds in heterotic theory to non--perturbative
vacua.\par In the Type I setting the special geometry is seen to
arise from a combined Born--Infeld Lagrangian and Chern--Simons
couplings, with the addition of suitable supersymmetry anomaly
counterterms. However in the $D=5$ context the special geometry is
simply dictated by the Chern--Simons five--dimensional coupling
\begin{eqnarray}
&&d_{ABC}\,\int A^A\wedge F^B\wedge F^C\,,
\end{eqnarray}
which specifically gives the following terms \cite{fms}
\begin{eqnarray}
&&Z\,dB\,dC\,\,;\,\,\,C\,v_{xy}\,dA^x\,
dA^y\,\,;\,\,\,B\,\tilde{v}_{xy}\,dA^x\, dA^y\,,
\end{eqnarray}
which determine uniquely the prepotential.\par For the case
$n_t=1$ the prepotential corresponds to a homogeneous symmetric
space \cite{adft}. Inspection of the prepotential for $n_t>1$
indicates that the space is homogeneous when the coefficients
$C_{rxy}$ satisfy some Clifford algebra relations and some further
relations between $n_t$ and $n_v$ hold true \cite{c,dwvp1,dwvp2}.
In this respect it would be interesting to see whether some
superstring models with $n_t>1$ exist with a homogeneous special
geometry , as in the $n_t=1$ case.
\section{Acknowledgements}
We would like to thank C. Angelantonj for enlightening
discussions. R.D. and M.T. would like to thank the Physics
Department of CERN, where part of this work was done, for its kind
hospitality. S.F. would like to thank the Physics Department of
Politecnico di Torino where part of this work was done, for its
kind hospitality.
\par
 The work of S.F. has been supported in
part by the D.O.E. grant DE-FG03-91ER40662, Task C, and in part by
the European Community's Human Potential Program under contract
HPRN-CT-2000-00131 Quantum Space-Time, in association with INFN
Frascati National Laboratories and R.D. and M. T. are associated
to Torino University.
\appendix
\section{Appendix. }
\label{appendiceA} In this appendix we compute the explicit
expression of the Riemann tensor for the dual K\"ahler manifold as
defined in \cite{dft} and discussed in section \ref{ale}.\par We
start from a very Special K\"ahler manifold of complex dimension
$n$, characterized by a cubic prepotential:
\begin{equation}
F(X)=\frac{1}{X^0}\,d_{abc}X^a X^b
X^c\,\,\,\,;\,\,\,\,\,\,a,b,c=1,\dots,n
\end{equation}
We denote be $\lambda^a=\Im z^a$, $z^a$ corresponding to the
special coordiate parametrization of the space. It is useful to
introduce the following quantities:
\begin{eqnarray}
\kappa(\lambda) &=& d_{abc}\lambda^a \lambda^b
\lambda^c\,\,\,;\,\,\,\,\kappa_a(\lambda)=d_{abc}\lambda^b
\lambda^c\,\,\,;\,\,\,\,\kappa_{ab}(\lambda)=d_{abc}
\lambda^c\,\,\,;\,\,\,\,V(\lambda)=\frac{\kappa(\lambda)}{6}
\end{eqnarray}
 The K\"ahler potential is:
 \begin{equation}
K(\lambda)=-\log (V(\lambda))
 \end{equation}
 The metric of the manifold and its inverse are therefore:
 \begin{eqnarray}
G_{a\bar{b}}&=&\frac{\partial^2 K}{\partial
\lambda^a\partial\lambda^b}=-6\,\left(\frac{\kappa_{ab}}{\kappa}-\frac{3}{2}\,
\frac{\kappa_a\kappa_b}{\kappa^2}\right)\nonumber\\
(G^{-1})^{a\bar{b}}&=&-\frac{1}{6}\,\left((\kappa^{-1})^{ab}\kappa-3\,
\lambda^a\lambda^b\right)
 \end{eqnarray}
The connection and the Riemann tensor are:
\begin{eqnarray}
\Gamma^c_{ab}&=&(G^{-1})^{c\bar{c}}\,\partial_a G_{\bar{c}
b}=-i\left(d_{ab\bar{c}}(\kappa^{-1})^{c\bar{c}}
-6\,\frac{\delta^c_{(a}
\kappa_{b)}}{\kappa}+3\frac{\kappa_{ab}\lambda^c}{\kappa}\right)\nonumber\\
R_a{}^d{}_b{}^c&=&-(G^{-1})^{d\bar{d}}\,\partial_{\bar{d}}\Gamma^c_{ab}=
-i(G^{-1})^{d
f}\,\frac{\partial}{\partial\lambda^f}\Gamma^c_{ab}=\nonumber\\
&&-2\delta^c_{(a}\delta^d_{b)}+d_{abf}\,(\kappa^{-1})^{f
(c}\,\lambda^{d)}+\frac{1}{2}\,(\kappa^{-1})^{cd}\,\kappa_{ab}-\frac{\kappa}{6}\,
(\kappa^{-1})^{d\bar{d}}(\kappa^{-1})^{ce}(\kappa^{-1})^{\bar{c}
f}d_{\bar{d}ef}\,d_{ab\bar{c}}\nonumber\\
\end{eqnarray}
I can be easily verified that the above expression coincides with
the general formula for the Riemann tensor in the Special
Geometry, namely
 \begin{equation}
R_a{}^d{}_b{}^c=-2\delta^c_{(a}\delta^d_{b)}+e^{2 K}\,
C_{abm}\,C^{cdm}
\end{equation}
if we identify $C_{abc}$ with $d_{abc}$.\par
 We now define the ``dual'' K\"ahler manifold by the following transformation of the
imaginary parts of the complex coordinates: $t_a=\kappa_a/2$. The
dual metric is
\begin{eqnarray}
g^{a\bar{b}}&=&\frac{36}{\kappa^2}\,(G^{-1})^{a\bar{b}}=\frac{\partial^2
\hat{K}}{\partial t_a\partial t_b}\nonumber\\
\hat{K}&=&2 K
\end{eqnarray}
The connection and the Riemann tensor for the dual manifold  are:
\begin{eqnarray}
\tilde{\Gamma}^{bc}_{d}&=&\frac{6}{\kappa}\,(G^{-1})^{cq}\,\Gamma^b_{qd}=
i\left(d_{qdf}(\kappa^{-1})^{cq}(\kappa^{-1})^{bf}
+\frac{6}{\kappa}\,\delta^{(b}_d
\lambda^{c)}-\frac{3}{\kappa}\,(\kappa^{-1})^{cb}\,\kappa_d
\right)\nonumber\\
\tilde{R}^c{}_a{}^b{}_d&=&-i(g^{-1})_{af}\,\frac{\partial}{\partial
t^f }\,\tilde{\Gamma}^{bc}_{d}=\nonumber\\
&&-\delta^c_{(a}\delta^b_{d)}-d_{fe(a}\,\kappa_{d)}(\kappa^{-1})^{cf}\,
(\kappa^{-1})^{be}+(\kappa^{-1})^{cb}\,\kappa_{ad}+\frac{\kappa}{3}\,
(\kappa^{-1})^{e(c}(\kappa^{-1})^{b)f}(\kappa^{-1})^{qp}d_{epa}\,d_{fqd}\nonumber\\
\end{eqnarray}
 We have verified, by means of computer aided computations,
that, for all \emph{homogeneous symmetric} $N=2$ manifolds
\begin{equation}
R_a{}^d{}_b{}^c=\tilde{R}_a{}^d{}_b{}^c\label{rr}
\end{equation}
which signals that the dual manifolds are of the same kind. Aside
from the homogeneous symmetric manifolds we have considered an
instance of non--homogeneous manifold defined by
\begin{equation}
V(\lambda)=d_1 (\lambda^1)^3+d_2 (\lambda^2)^3
\end{equation}
and an instance of homogeneous non--symmetric manifold, namely
$K(1,1)$. In both cases equality (\ref{rr}) does not hold and
therefore the two spaces are necessarily different.

\end{document}